\documentclass[useAMS,usenatbib]{mn2e}

\usepackage{times}

\usepackage[dvips]{graphicx}
\usepackage{amsmath,amssymb}
\usepackage{bm}

\title[Star Formation Triggered by Supernova Explosions 
in Young Galaxies]{Star Formation Triggered by Supernova Explosions 
in Young Galaxies} 
\author[T. Nagakura, T. Hosokawa, K. Omukai]{Takanori
Nagakura$^{1}$\thanks{E-mail: 
nagakura@th.nao.ac.jp (TN) ; hosokawa@th.nao.ac.jp (TH) ;
omukai@th.nao.ac.jp (KO)} 
, Takashi Hosokawa$^{1}$, and Kazuyuki Omukai$^{1}$\\
$^{1}$National Astronomical Observatory of Japan, Osawa, Mitaka, 
Tokyo 181-8588} 
\begin{document}


\pagerange{\pageref{firstpage}--\pageref{lastpage}} \pubyear{}

\maketitle

\label{firstpage}

\begin{abstract}
 We study the evolution of supernova remnants in a low-metallicity 
 medium $Z/Z_{\odot} = 10^{-4}$ -- $10^{-2}$ in the early universe, using
 one-dimensional hydrodynamics with non-equilibrium chemistry.
 Once a post-shock layer is able to cool radiatively, 
 a dense shell forms behind the shock.
 If this shell becomes gravitationally unstable and fragments into pieces,
 next-generation stars are expected to form from these fragments.
 To explore the possibility of this triggered star formation, 
 we apply a linear perturbation analysis of an expanding shell to our
 results and constrain the parameter range of ambient density, explosion
 energy, and metallicity where fragmentation of the shell occurs.
 For the explosion energy of $10^{51}{\rm ergs} (10^{52}{\rm ergs})$, 
 the shell fragmentation occurs for ambient densities higher than 
 $\gtrsim 10^{2} {\rm cm^{-3}}$ ($10$ ${\rm cm^{-3}}$, respectively). 
 This condition depends little on the metallicity in the ranges we examined.
 We find that the mode of star formation triggered occurs 
 only in massive ($\gtrsim 10^{8}M_{\odot}$) haloes.
\end{abstract}

\begin{keywords}
 cosmology:theory -- galaxies:formation -- high-redshift --
 ISM:supernova remnants -- shock waves -- stars:formation
\end{keywords}

\section{Introduction}
\label{sec:intro}

Understanding of star formation and its feedback effects in the 
high-redshift universe is a key to unravelling how the
primeval universe has evolved into the variety of luminous objects we
observe today.
Recent numerical studies have indicated that the first stars, or Pop
III.1 stars (O'Shea et al. 2008), presumably have typical mass of
$\gtrsim 100$ M$_{\odot}$ (Abel, Bryan \& Norman 2002; Bromm, Coppi \&
Larson 2002; Omukai \& Palla 2003; Yoshida et al. 2006).
Copious amounts of radiation from such very massive stars photo-ionize
hydrogen atoms and photo-dissociate molecular hydrogen in their
surroundings (Kitayama et al. 2004; Whalen et al. 2004).  
As well as the radiative feedback, the first star deaths have an impact
on subsequent star formation around them.  
The ultimate fate of metal-free stars is determined by their mass.
Stellar evolution models predict that 
stars with $40$ -- $140$ M$_{\odot}$ finally collapse into black
holes without supernovae, while those with $10$ -- $40$
M$_{\odot}$ or $140$ -- $260$ M$_{\odot}$ explode as Type II supernovae 
or pair-instability supernovae (PISNe), respectively (Heger \& Woosley
2002; Umeda \& Nomoto 2002).

If the first star dies without supernova, the H{\sc ii} region is left over
around a remnant black hole.
At the centre of the fossil H{\sc ii} region, H$_2$ is quickly replenished 
owing to a large amount of electrons, which are the catalysts for the
H$_2$ formation reaction.  
The resultant fast radiative cooling allows a small portion 
of the gas to re-collapse and turn into the next-generation stars, 
provided that no external ultraviolet (UV) radiation 
effectively dissociates H$_2$ and the dark halo mass is higher than
$10^5$ M$_{\odot}$ (Nagakura \& Omukai 2005).
In addition, deuterated hydrogen (HD) molecules abundantly form in such
regions and further reduce the gas temperature to that of the CMB. 
The re-collapsing gas fragments with typical mass 
of $\simeq 40$ M$_{\odot}$ (Yoshida et al. 2007), which is  
an order of magnitude lower than that of the first
stars (see also, Uehara \& Inutsuka 2000; Nakamura \& Umemura 2002).

If the first star explodes as a supernova, the surrounding medium is
severely affected by the blast wave. 
The blast wave removes most of the gas from a low-mass
halo ($\lesssim 10^6$ M$_{\odot}$) even for the explosion energy 
as low as $10^{50}$ erg (Kitayama \& Yoshida 2005; Whalen et al. 2008),
because the gas density has been decreased to $0.1$ -- $1$ cm$^{-3}$
inside the H{\sc ii} region. 
Evacuation of the gas from the halo quenches subsequent star 
formation in the same halo without a substantial gas supply through 
merging with other haloes (Greif et al. 2007). 
For a supernova in higher-mass haloes ($\gtrsim 10^7$ M$_{\odot}$),
the situations are quite different.
The halo gas remains in the H{\sc ii} region owing to a deeper dark matter
potential well (Kitayama et al. 2004).
The blast wave stagnates inside the halo and
sweeps a considerable amount of gas into a dense shell (Whalen et al. 2008).
Such a shell is a potential site for next-generation star formation.
The stability analyses of linear perturbations on an expanding and
decelerating thin shell predict that small density perturbations 
can grow to trigger gravitational instabilities, provided that 
the temperature in the shell becomes low enough and 
sufficient gas is taken into the shell (Elmegreen 1994; Whitworth et
al. 1994). 
Some authors have investigated the possibility of star formation triggered
by the first supernova explosions (Ferrara 1998; Mackey, Bromm \&
Hernquist 2003; Bromm, Yoshida \& Hernquist 2003; Salvaterra, Ferrara \&
Schneider 2004; Machida et al. 2005; Vasiliev, Vorobyov \& Shchekinov
2008; Whalen et al. 2008).  
Applying a classical supernova remnant (SNR) model to a primordial gas,
Salvaterra et al. (2004) and Machida et al. (2005) found the 
combination of the explosion energy and the ambient density 
where the shell fragmentation occurs, 
and discussed the typical mass of the stars forming in the shell. 
However, their results are somewhat different 
because of different one-zone modelling.
To solve this uncertainty, more detailed calculation is needed.

Supernovae also disperse heavy elements into the intergalactic medium
(IGM) (e.g., Kitayama \& Yoshida 2005).
Greif et al. (2007) showed that the metals are preferentially
transfered into voids, while some of them collide with neighbouring 
haloes and mix with the gas in them through hydrodynamical 
instabilities such as Kelvin-Helmholtz instability.
Cen \& Riquelme (2008) showed that the mixing occurs only at the edge of 
haloes of $10^{6-7}~M_\odot$ at $z = 10$.
However, irradiation of ionizing photons from neighbouring haloes, 
which was omitted from their analysis, may enhance the metal
enrichment of lower-mass haloes at higher redshifts. 
Turbulent motions excited by hierarchical mergers also causes
the efficient metal mixing (Wise \& Abel 2007; Grief et al. 2008).  
Metals contribute to the radiative cooling of gas, 
having an effect on dynamical evolution of SNRs.
Thornton et al. (1998) investigated this issue for a wide 
metallicity range with the intention of including the SN feedback 
in galaxy-formation simulations. 
They showed that the metal cooling modifies the redistribution of
supernova energy to the surroundings  
for metallicity $>0.01 Z_{\odot}$.
However, they paid no attention to the structure and evolution of the
supernova-driven shell nor to the possibility of subsequent fragmentation.

The aim of this paper is to find the condition where the supernova-driven 
shell fragments and triggers subsequent star formation
in young galaxies.
Towards this end, we study the evolution of supernova remnants 
in the low-metallicity interstellar medium by way of 
hydrodynamics with spherical symmetry.
This paper is organized as follows:
In Section \ref{sec:models}, we describe the 
numerical method and input physics. 
The results are presented in Section \ref{sec:results}, where
we first describe dynamical evolution of an SNR and 
the structure of the shell, focusing on the roles of metals, 
and then discuss the possibility of the shell fragmentation using
the linear stability analysis by Elmegreen (1994).
In Section \ref{sec:discuss}, by using the above result, 
we constrain the mass range of haloes where the triggered 
star formation occurs.
In addition, we discuss effects of processes not included in our analysis.
Finally, we summarize our study in Section \ref{sec:summary}.
Throughout this paper, we adopt a flat $\Lambda$CDM model with
parameters $\Omega_{\Lambda}=0.73$, $\Omega_{m}=0.27$, 
$\Omega_b=0.044$, and Hubble parameter $h=0.71$ (Spergel et al. 2003).

\section[]{The Model}
\label{sec:models}

\subsection{Hydrodynamics and Chemistry}

We solve the following basis equations for 
one-dimensional hydrodynamics with spherical symmetry: 
\begin{align}
& \frac{\partial \rho}{\partial t} + \frac{1}{r^2}
 \frac{\partial}{\partial r} (\rho v) = 0, \label{eq:ce} \\
& \frac{\partial v}{\partial t} + v \frac{\partial v}{\partial r} = -
 \frac{1}{\rho} \frac{\partial p}{\partial r} - \frac{G m(r)}{r^2} ,
 \label{eq:eom} \\
& \frac{\partial}{\partial t}(\rho e) + \frac{1}{r^2}
 \frac{\partial}{\partial r} (r^2 \rho e v) + \frac{p}{r^2}
 \frac{\partial}{\partial r}(r^2 v) \notag \\
& \quad \quad = - \Lambda(\rho,T) + \frac{1}{r^2}
 \frac{\partial}{\partial r} \left[ \kappa (T) r^2 \frac{\partial
 T}{\partial r} \right] , \label{eq:ee} \\
& p = (\gamma - 1) \rho e 
 = \frac{k_{\rm B} T}{\mu m_{\rm H}} \rho , \label{eq:eos}
\end{align}
where $\rho$ is the mass density of the gas, $v$ the velocity of the
fluid elements, $p$ the pressure, $e$ the internal energy per unit
mass, $m(r)$ the enclosed mass within the radius $r$, $\gamma$ the 
adiabatic exponent, $\mu$ the mean molecular weight, $m_{\rm H}$ the
mass of a hydrogen atom, $\Lambda$ the 
radiative cooling rate per unit volume and time, $k_{\rm B}$ the
Boltzmann's constant, and $\kappa(T)$ the thermal conductivity.
The thermal conductivity is the sum of contributions from atomic
diffusion at low temperatures (Parker 1953) 
and electronic diffusion at high temperatures (Spitzer 1962; Cowie \&
McKee 1977).  
In our calculations, a cooling layer behind a shock driven by a
supernova becomes thermally unstable for wavelengths shorter than a
so-called Field length (Field 1965) : 
$\lambda_F = (\kappa T/\Lambda)^{1/2}$.
Thermal conduction tends to curtail the thermal instability, but the Field
length is many orders of magnitude shorter than a characteristic length
(e.g., a radius of an SNR).  
The limitation of our simulation resources makes it difficult to resolve
the Field length. 
To suppress unphysical oscillations due to the thermal instability
occurring below the spatial resolution limit, we temporarily enhance the
thermal conductivity only for the thermally unstable cooling layer at
the incipient stage of the shell formation (see Hosokawa \& Inutsuka 2006).  
This prescription smoothes the temperature structure out on the 
newly defined Field length, but does not significantly alter the thermal
evolution of the shell because the Field length is still shorter than 
the scale-length of temperature variation.

The hydrodynamics are solved with a second-order Godunov
method in the Lagrangian coordinate (van Leer 1979). 
We split the gas into fluid elements which are unequal in mass such that
more fluid elements exist in the post-shock regions. 
We set the number of the total fluid elements to 1000 and choose a
simulation box of length $100$ -- $1000$ pc, depending on the explosion
energy and the ambient density.
We have confirmed that the results do not change when the total number
of mesh points is doubled. 
To check the accuracy of our code,
we have carried out a test simulation on a dynamical expansion of a
point-source explosion without any dissipative processes and ascertained
that the result is excellently consistent 
with the Sedov-Taylor solution.

Along with the hydrodynamics, we solve a set of
kinetic equations for chemical species.
We include 15 primordial components and 4 heavy elements as follows : 
${\rm H}$, ${\rm H}^{+}$, ${\rm H}^{-}$,  
${\rm H}_2$, ${\rm H}_2^{+}$, ${\rm HeH}^{+}$, ${\rm He}$, 
${\rm He}^{+}$, ${\rm He}^{++}$, 
$e^{-}$, ${\rm D}$, ${\rm D}^{+}$, ${\rm D}^{-}$, ${\rm HD}$, 
${\rm HD}^{+}$, ${\rm C}$, ${\rm C}^{+}$, ${\rm O}$, and ${\rm O}^{+}$.
The kinetic equation for the $i$-th species is  
\begin{align}
\frac{dy(i)}{dt} 
& = - n_{\rm H} y(i) \sum_{j} k_{ij} y(j) 
+ n_{\rm H} \sum_{lm} k_{lm} y(l) y(m) \notag \\
& + n_{\rm H}^2 \sum_{ijk} k_{ijk} y(i) y(j) y(k)  
, \label{eq:chem_eq}
\end{align}
where $n_{\rm H}$ is the number density of hydrogen nuclei, $y(i)$ 
the fraction of the number density for the $i$-th chemical species to 
$n_{\rm H}$, $k_{ij}$ and $k_{lm}$ the destruction and the formation
rate of $y(i)$, respectively, and $k_{ijk}$ the reaction rates for
three-body reactions. 
We include 53 reactions among these species on the basis of the
minimum model suggested by Omukai et
al. (2005) and listed in Table \ref{tab:reactions}.
The primordial chemical networks are based on Abel et
al. (1997) and Galli \& Palla (1998).
The reactions involved with carbon and oxygen are the recombination of
C$^{+}$ (reaction [R42]) and O$^{+}$ (reaction [R43]).  
In addition, we incorporate the charge exchange reactions between H and
O$^{+}$ (reaction [R44]), and between H$^{+}$ and O (reaction [R45]) 
because these reactions occur quickly due to a slight difference
among their ionization potentials. 
We do not take into account the charge exchange reactions between C and
H$^{+}$ or between H and C$^{+}$ because these reactions are much less
important than reactions [R44] and [R45] (Glover \& Jappsen 2007).
Although the emissions from the compounds of carbon or oxygen such as OH,
and H$_2$O slightly contribute to the total radiative cooling at densities 
ranging from $10^5$ cm$^{-3}$ to $10^{10}$ cm$^{-3}$ (see, Omukai et
al. 2005), we omit the reactions among them because they have 
only minor effects in the density range we explored.
We apply the set of equations to each fluid element and solve them with
an implicit scheme because of the stiffness of these equations.
The adopted time steps are actually shorter than the chemical time.
We implicity solve the chemical reactions to make sure the 
numerical stability. We have confirmed that our results do not change
even with the shorter time steps.

\subsection{Radiative Processes}

The net cooling rate $\Lambda$ in equation (\ref{eq:ee}) is the sum of
the contributions from radiative cooling processes.
In simulating supernova explosions, we need to treat both the hot interior
bubble and dense shell driven by the blast wave.
The temperature range is wide, spanning a wide range from $10$ K to
$10^{9}$ K.  
In the high-temperature regime of $>10^4$ K, 
the important processes include thermal bremsstrahlung emission, 
inverse Compton scattering of the CMB, collisional ionization 
and excitation of hydrogen and helium atoms, recombination of 
hydrogen and helium.
Their cooling rates are taken from Cen (1992).
On the other hand, at lower temperatures, the radiative processes by
molecules and metals become important.
We take account of H$_2$ and HD ro-vibrational cooling 
(Hollenbach \& McKee 1979; Galli \& Palla 1998).
Recently, Glover \& Abel (2008) have computed new H$_2$ cooling rates
for e$^-$ -- H$_2$ and H$^+$ -- H$_2$ collision channels in addition to
the rate we included. 
We do not include these updated rates for simplicity. 
This effect might slightly reduce temperature of the post-shock gas,
but not affect our conclusions.
We also include the meta-stable and fine-structure line cooling by 
carbon and oxygen (Hollenbach \& McKee 1989): 
 [C{\sc i}]$\lambda 9823$, $\lambda 4622$, and $\lambda 8727$ 
; [C{\sc ii}]$\lambda 2326$ 
; [O{\sc i}]$\lambda 6300$, $\lambda 2972$, and $\lambda 5577$  
; [O{\sc ii}]$\lambda \lambda 3729, 3726$, and 508 $\mu$m 
; [C{\sc i}]$609.2 \mu$m, $229.9 \mu$m, and $369.0\mu$m 
; [O{\sc i}]$63.1 \mu$m, $44.2 \mu$m, and $145.6 \mu$m.
The cooling rates are derived by solving the statistical equilibrium
between the energy levels. 
We omit the metal line cooling at temperatures higher than
10$^4$ K because its contribution is less important in the metallicity
range we consider (see, Sutherland \& Dopita 1993).

Some metals are expected to have condensed into dust grains 
even in the early universe (Nozawa et al. 2003; Bianchi \& Schneider
2007). 
The dust can be an efficient radiator, but it is important only 
at densities far higher than we consider here.
Also, it plays a role in a heating source through
photo-electric heating under a strong radiation field.
This rate depends on the properties of the 
radiation field as well as dust grains, both of which are 
quite uncertain in the high-redshift universe.
Here, we just omit the dust effect for simplicity.

The CMB temperature is quite a high value of $\sim 44$ K at $z\sim 15$.
If a gas temperature decreases below that of the CMB, the CMB
photons heat the gas through radiative pumping.
This effect is taken into consideration in our calculation.
We do not include other heating such as cosmic rays or 
external radiation from other sources.

\subsection{Initial Condition}
\label{sec:condition}

Before the supernova explosion,
the progenitor star ionizes the surrounding medium. 
Dynamical expansion of the H{\sc ii} region has
been investigated by several authors (Whalen et al. 2004;
Kitayama et al. 2004; Hosokawa \& Inutsuka 2006).
A shock wave associated with the ionization front pushes 
a large amount of the circumstellar gas away, 
leveling the density off.  
Thus, we simply assume that the surrounding gas is homogeneous and 
fully ionized at temperature $10^4$ K.
In reality, the typical temperature in H{\sc ii} regions 
around Population III stars is somewhat higher than $10^4$ K 
(Whalen et al. 2004; Kitayama et al. 2004).
However, this difference will not be so significant in our calculations.
In an early phase of the expansion, ambient thermal pressure is negligible 
compared with ram pressure exerted on the shell. Temperature in the
fossil H{\sc ii} regions quickly falls below 10$^4$ K by the Lyman-$\alpha$ cooling.
The initial temperature difference will be wiped out before the ambient 
thermal pressure acts to decelerate the shell.
The ambient density depends on the conditions of 
formation sites of the progenitors.
Here, we treat it as a free parameter and study 
the cases of $n_0=0.1, 1, 10$, and $10^3$ cm$^{-3}$. 
We start the simulations at redshift $z=15$.

We inject the supernova energy $E_{\rm SN}$ in the form of thermal energy 
and load the ejecta mass into several central meshes.
Since the explosion energy can differ from event to event,
we studied the three cases of $E_{\rm SN}=10^{51}$, $10^{52}$ and
$10^{53}$ ergs, which correspond to normal core-collapse
supernovae, hypernovae and PISNe, respectively.
Note that we performed the PISN runs only for low-metallicity 
($Z=10^{-4} Z_{\odot}$) because metal enrichment reduces a typical
fragment scale of a collapsing gas, and hence stellar mass (e.g., Omukai
et al. 2005). 
The result depends little on the ejecta mass 
because it is much smaller than the mass swept up later into the shell. 
Hence, we simply fix it to $10$ M$_{\odot}$.

For the elemental abundances in ISM, the number fractions of He and D nuclei
to the hydrogen ones are $7.9 \times 10^{-2}$ and $2.5 \times 10^{-5}$
respectively. 
We study the cases with metallicities 
$Z/Z_{\odot}=10^{-4},10^{-3},10^{-2}$.
For simplicity, we assume the metal abundance ratios
are in proportion to the solar values.
The number fractions of carbon and oxygen are $3.6 \times 10^{-4}$ and
$8.5 \times$ $10^{-4}$, respectively, for the solar metallicity 
(Anders \& Grevesse 1989).

\section{Results}
\label{sec:results}

\subsection{Dynamical Expansion of a Supernova Remnant}

\begin{figure*}
\begin{center}
 \scalebox{1.0}{
  \includegraphics*[width=\hsize]{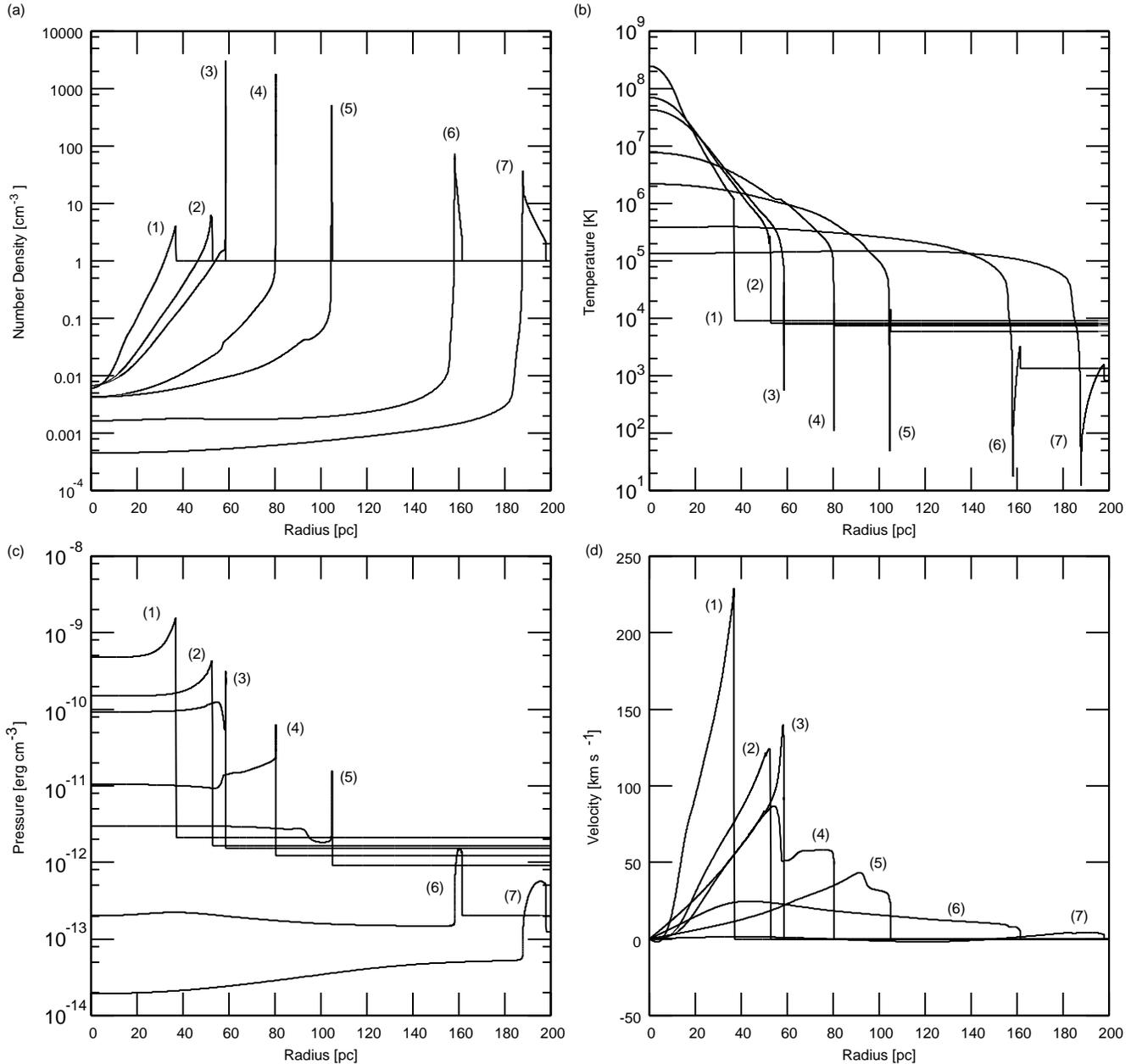}}
\end{center}
 \caption{Evolution of radial distributions of 
(a) density, (b) temperature, (c) pressure,
 and (d) velocity for the fiducial run 
 (E$_{\rm SN}=10^{52}$ erg, n$_0=1$ cm$^{-3}$, $Z=10^{-4} Z_{\odot}$).
 Shown are the distributions at 
 (1) $9.0 \times 10^{3}$, (2) $4.8 \times 10^{4}$, (3) $1.7 \times 10^{5}$, 
 (4) $4.9 \times 10^{5}$, (5) $1.1 \times 10^{6}$, (6) $5.0 \times 10^6$, 
 and (7) $1.0 \times 10^{7}$ yr, respectively. The epochs (1) and
 (2) correspond to the Sedov-Taylor phase, and those from (3) to (7) 
 to the snowplough phase.}
 \label{fig1}
\end{figure*}

We first present the result for the parameter set of 
$E_{\rm SN}=10^{52}$ erg, $n_0=1$ cm$^{-3}$, and $Z=10^{-4}Z_{\odot}$,
which we hereafter call the fiducial run. 
Fig. \ref{fig1} shows the snapshots of (a) the density,
(b) temperature, (c) pressure, and (d) velocity distribution as a
function of the radius.
The numbers in Fig. \ref{fig1} represent the structures 
at time 
(1) $9.0 \times 10^3$, (2) $4.8 \times 10^4$, (3) $1.7 \times 10^5$,
(4) $4.9 \times 10^5$, (5) $1.1 \times 10^6$, (6) $5.0 \times 10^6$,
and (7) $1.0 \times 10^7$ yr. 
A contact discontinuity forms quite early at the boundary between the
stellar remnants and the surrounding medium, which is
consistent with the previous studies (Chevalier 1974; Cioffi, McKee \&
Bertschinger 1988; Thornton et al. 1998).  
However, this discontinuity disappears before the state (1) because the
heat transfer causes the gas slightly outside it to flow into a hot
interior bubble while smoothing out the boundary.  
The strong shock evolves adiabatically until $t=8.5 \times 10^4$ yr.
The post-shock density is $(\gamma+1)/(\gamma-1)=4$ times as high as
that of the ambient medium.  
The shock front moves outwards in conformity with the canonical
power-law form ; $R_{\rm sh} \propto t^{\eta}$, where the exponent
$\eta=2/5$ during the Sedov-Taylor (ST) phase.

The expansion time of the shock front, $R_{\rm sh}/\dot{R}_{\rm sh}$, 
increases linearly with the time elapsed during the ST phase, while the
radiative cooling time in the post-shock layer, $\rho e/\Lambda$, 
continues to decrease. 
After the cooling time becomes shorter than the 
expansion time at  $t=8.5 \times 10^4$ yr,
the post-shock temperature quickly plunges to a low value
and the gas in the post-shock layer is compressed to a 
thin shell.
The mean pressure in the hot bubble is several orders of
magnitude higher than that of the ambient medium at this moment
(Fig. \ref{fig1} c),
and drives the outward motion of the shell.
This is called the pressure-driven snowplough (PDS) phase.
An analytic model of a point-source explosion exhibits $\eta=2/7$
in the absence of radiative cooling in the hot bubble (McKee \& Ostriker
1977).
The power-law index of our results, however, deviates from that 
of the analytic model $2/7$ because the interior pressure at the ST phase
slightly increases the effective value of $\eta$
(Cioffi et al. 1988; see Whalen et al. 2008, in detail).
Note that the expansion law of the bubble will change with
more realistic density distribution in the relic HII region, 
e.g., clumpiness, and radial density gradient.
These effects should be separately examined in future studies.
This phase continues as long as the mean interior pressure is much higher
than the ambient pressure. 

Radiative cooling is inefficient inside the bubble because of 
the extremely low density, and the gas loses thermal energy 
through adiabatic expansion instead. 
The mean interior pressure continues to decrease and becomes
comparable to the ambient pressure at $t=4.6$ Myr.
The motion of the shell is then driven by its momentum.
This is a so-called momentum-conserving snowplough (MCS) phase. 
The thin shell model predicts that $\eta=1/4$, but it is slightly higher
in our results because of the contribution  
of finite pressure in the bubble (Cioffi et al. 1988).
The shell's expansion velocity eventually falls to the sound speed
of the ambient medium.
In reality, the shell loses its identity thereafter and mixes
with the ISM. 
This is the end of the lifetime of the shell.

The results of the other runs are similar to the fiducial one.
In all cases, the SNRs enter the MCS phase 
before the end of the shell lifetime.
This is because the ambient temperature monotonically decreases to a low
value by radiative cooling.
If the ambient matter is heated, for example, by 
external radiation, the MCS phase is unlikely to be reached.

\begin{figure*}
\begin{center}
 \scalebox{1.0}{
  \includegraphics*[width=\hsize]{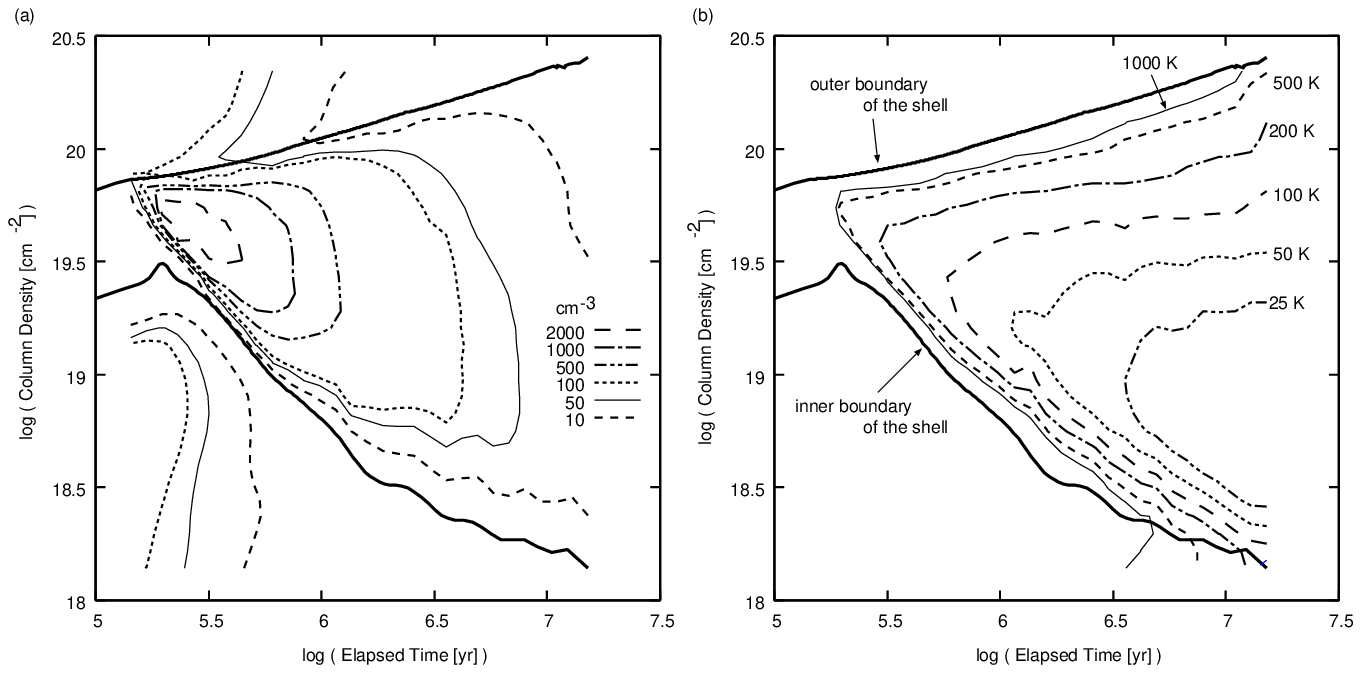}}
\end{center}
 \caption{Contour maps for densities (a) and temperatures (b) in the
 shell as a function of time for the fiducial run 
 (E$_{\rm SN}=10^{52}$ erg, n$_0=1$ cm$^{-3}$,  
 $Z=10^{-4} Z_{\odot}$).
 The longitudinal axis is the column density measured from the centre.
 The two thickest lines represent the boundaries of the shell. 
 In figure (a), we show the iso-density surfaces of 2000 cm$^{-3}$ (dashed),
 1000 cm$^{-3}$ (dotted-and-dashed), 500 cm$^{-3}$
 (double-dotted-and-dashed), 100 cm$^{-3}$ (dotted),
 50 cm$^{-3}$ (solid), and 10 cm$^{-3}$ (thick-dashed line). 
 Figure (b) represents the iso-temperature surfaces of 1000 K (solid),
 500 K (dashed), 200 K (dotted-and-dashed), 100 K (long-dashed), 50 K
 (dotted), and 25 K (double-dotted-and-dashed line) .}
 \label{fig2}
\end{figure*}%

\subsection{Thermal Evolution and Internal Structure of the Shell}
\label{result2}

\begin{figure}
\begin{center}
 \scalebox{1.0}{
  \includegraphics*[width=\hsize]{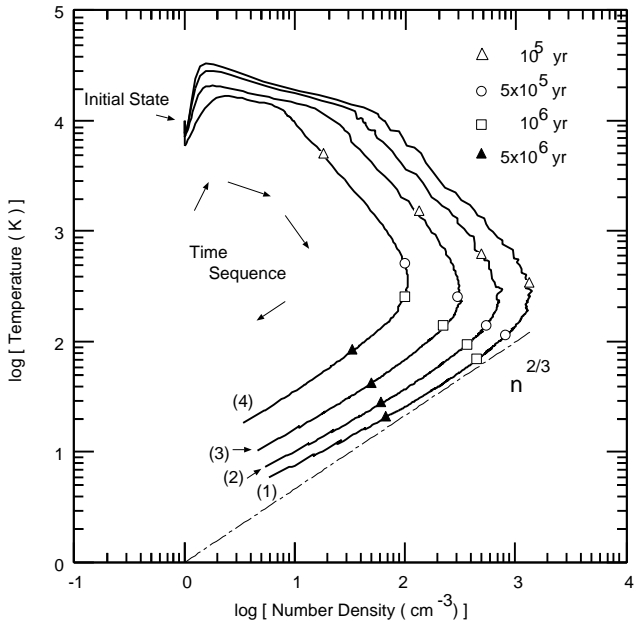}}
\end{center}
 \caption{The evolution of four fluid elements
 on the density-temperature diagram in the fiducial run. 
 The initial locations of the fluid elements are 
 (1) 68, (2) 77, (3) 88, and (4) 104 pc from the centre. 
 Those elements enter the shell in the snowplough phase. 
 The arrows indicate the direction of evolution. The circles, triangles,
 squares and black triangles are indicate the times $10^5$ yr, 
 $5 \times 10^5$ yr, $10^6$ yr and $5 \times 106$ yr, respectively,
 after which the shock-compressed gas cools to $10^4$ K.} 
 \label{fig3}
\end{figure}

\begin{figure}
\begin{center}
 \scalebox{1.0}{
  \includegraphics*[width=\hsize]{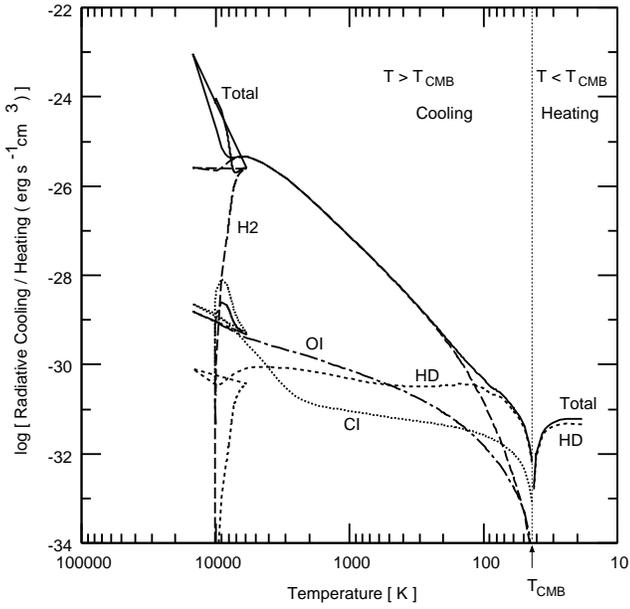}}
\end{center}
 \caption{Radiative cooling and heating rate of 
 a typical fluid element (4 in Figure \ref{fig3}) against
 the temperature. The vertical dotted line shows the CMB temperature, 
 $T_{\rm CMB}=44$ K, below which the processes work as heating. 
 The lines represent the radiative rates of H$_2$ (long-dashed),
 HD (dashed), C{\sc i} (dotted) and O{\sc i} (dotted-and-dashed). 
 The solid line shows the total rate, including the H atomic cooling.}
 \label{fig4}
\end{figure}

\begin{figure}
\begin{center}
 \scalebox{1.0}{
  \includegraphics*[width=\hsize]{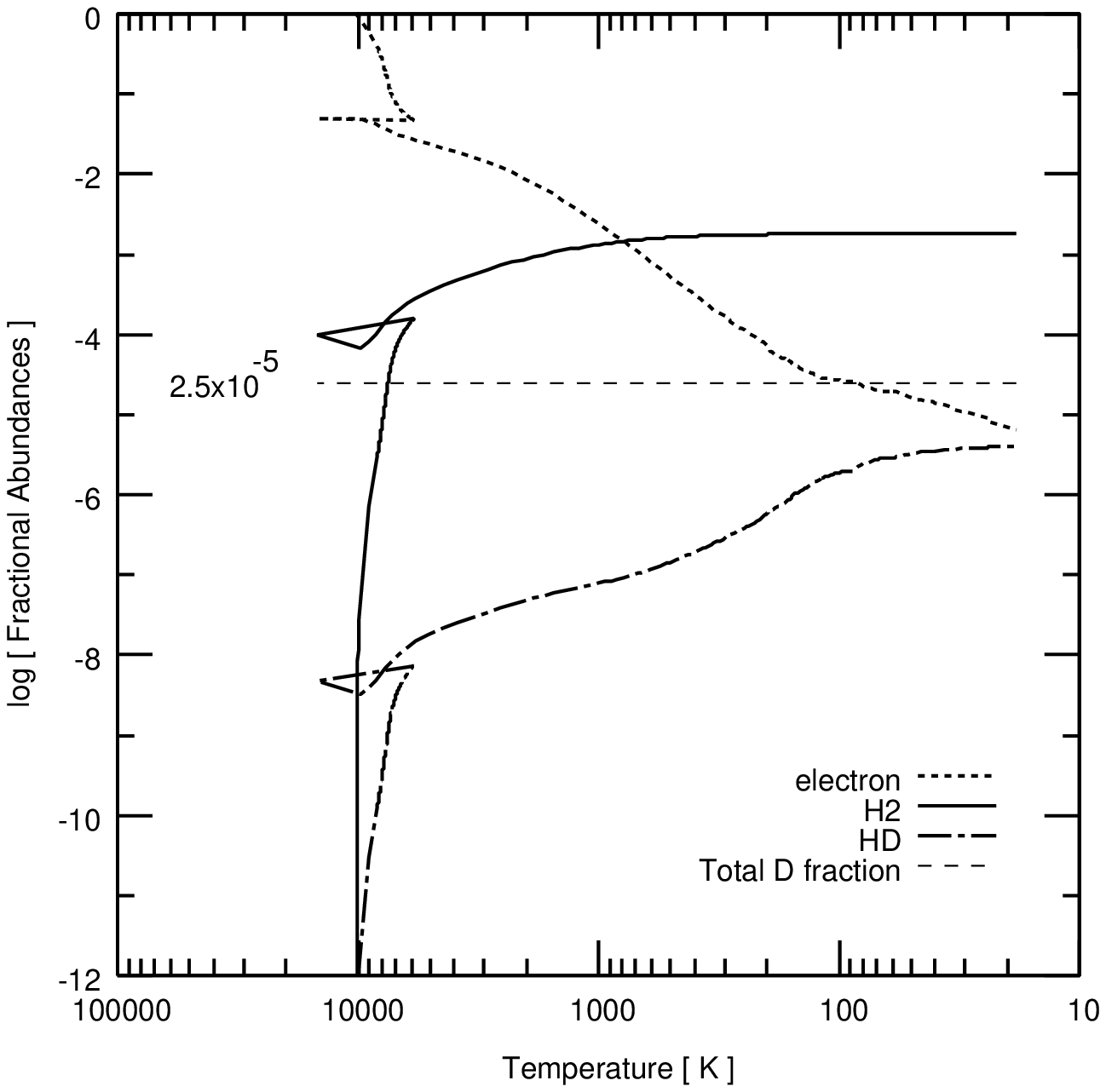}}
\end{center}
 \caption{Evolution of the H$_2$ (solid), HD (dotted-and-dashed) and 
 electron (dotted) fractions 
 at the same fluid element as Figure \ref{fig4} against the
 temperature. 
 The horizontal dashed-line represents the total deuterium abundance 
 of $2.5 \times 10^{-5}$. 
 The effect of the shock is observed as the kinks in the fraction 
 curves.
 Thereafter the temperature continues to decrease, 
 while the H$_2$, and HD fractions gradually increase.} 
 \label{fig5}
\end{figure}

Fig. \ref{fig2} shows time evolution of the internal structure of the shell. 
Radial density and temperature profiles are presented against
column density of hydrogen nuclei measured from the center. 
The column density of the shell increases as the shock front sweeps 
up the ambient materials, whereas the column density of the bubble
decreases with time, because the mean density in the bubble quickly 
falls by the expansion.
The mean density in the shell gradually decreases over the snowplough 
phase. This is because the shock strength becomes weaker and weaker 
as the bubble expands. The minimum temperature in the shell 
decreases with time. Note that it eventually falls below the CMB value. 
To see the physical processes, we plot in Fig. \ref{fig3} the
evolutionary trajectories of some typical fluid elements on the
temperature-density diagram. 
These elements are taken into the shell in the snowplough phase.
After they slightly drop from the initial location at
$(n_0,E_{\rm SN})=(1~ {\rm cm^{-3}}, 10^{4}~{\rm K})$ by radiative cooling, 
the temperatures suddenly rise with the shock arrival.
Due to fast cooling from atomic hydrogen, 
the post-shock layer cannot be resolved until 
the temperature falls back to $\sim 10^{4}~{\rm K}$, where 
the post-shock pressure balances with the ram pressure.
However, our insufficient resolution there does not affect 
the evolution thereafter because the time-scale for the atomic cooling
is very small compared with the entire evolutionary time-scale of the shell.
For element (4) in Fig. \ref{fig3},
the radiative cooling rate and the chemical fractions 
are shown in Figures \ref{fig4} and \ref{fig5},
respectively.
Following the shock heating, the gas cools to somewhat 
below $10^4$ K through the Ly $\alpha$ emission.
By this time, abundant H$_2$ has formed as a result of 
a large amount of available electrons. 
The H$_2$ abundance reaches as high as $\sim 10^{-3}$
(Fig. \ref{fig5}), about an order of magnitude higher
than that formed in an almost neutral gas.
Below $\simeq$ 8000 K, the H$_2$ becomes the most important 
coolant (Fig. \ref{fig4}).
The H$_2$ cooling reduces the temperature down to $\sim 200$ K,
giving rise to compression of the gas. 
The metal line cooling has a negligible effect during this phase  
because of the low abundance of metals (Fig. \ref{fig4}). 
Below $\simeq 200$ K, the H$_2$ cooling rate significantly drops and the
other coolants cannot compensate for the inefficiency.
At this moment, radiative cooling cannot reduce the shell pressure
rapidly, while the ram pressure decreases faster than the shell pressure.
As a result, the shell pressure becomes higher than the ram pressure
and the shell begins to gradually distend to maintain the balance of pressure. 
This is seen in Fig. \ref{fig3} as the turn-around of 
the trajectories at $\sim 200$ K.
The adiabatic expansion phase continues for 
$\sim 10^6$ -- $10^7$ yr, which is much greater than a typical
time-scale for the H$_2$ cooling phase ($\sim 10^5$ -- $10^6$ yr), 
until the shell dissolves into the ambient medium.

At low temperatures ($\lesssim 150$ K),
HD is known to form abundantly via the rightward reaction of  
\begin{equation}
{\rm D}^{+} + {\rm H}_2 \rightleftarrows {\rm H}^{+} + {\rm HD}.
\label{eq:HDfd2}
\end{equation}
Although the temperature falls below the threshold for 
HD formation in our case, the density within the shell decreases as well.
The reaction \label{eq:HDfd2} thus proceeds slower than the evolutionary
time-scale of the shell, and the HD fraction is frozen to the final
abundance of $\sim 10^{-6}$, which is an order of magnitude smaller 
than the total deuterium abundance (Fig. \ref{fig5}).
The gradual density decrease causes the HD cooling to be inefficient. 
For $T<T_{\rm CMB}$, the CMB heating via HD lines is weak 
and the trajectories in Fig. \ref{fig3} exhibit the
relation for the adiabatic expansion, $T \propto n^{2/3}$,
except the highest density case of trajectory 1.

Note that the cooling time in the post-shock gas lengthens  
for smaller shock velocity, and reaches $\sim 10^6$ yr for 
a gas entering the shell during the MCS phase.
Thus, the temperature at the outer edge of the shell remains high
(Fig. \ref{fig2} b).
We use this result to explore the conditions for the shell
fragmentation in Section \ref{sec:fragment}. 
 

\subsection{Metallicity Effects on the Shell Evolution}

\begin{figure}
\begin{center}
 \scalebox{1.0}{
  \includegraphics*[width=\hsize]{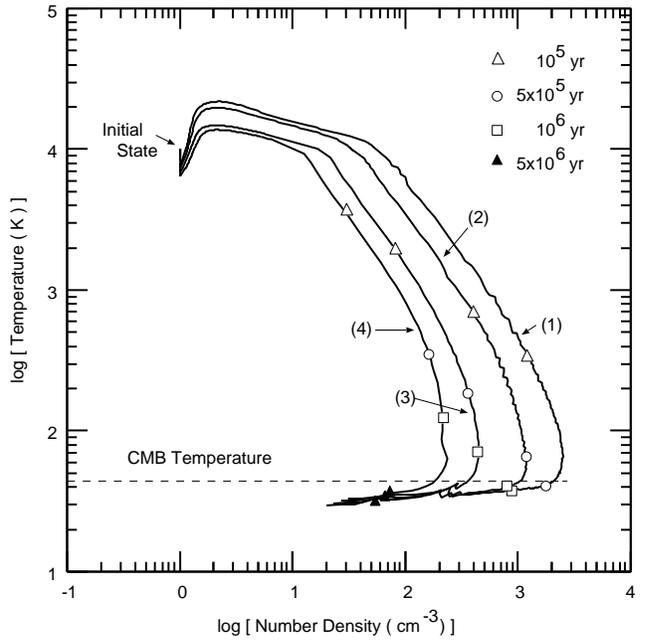}}
\end{center}
 \caption{Same as Fig. \ref{fig3}, except for the metallicity
 $Z=10^{-2} Z_{\odot}$. The horizontal dashed-line represents the 
 CMB temperature at $z=15$.}
 \label{fig6}
\end{figure}

\begin{figure}
\begin{center}
 \scalebox{1.0}{
  \includegraphics*[width=\hsize]{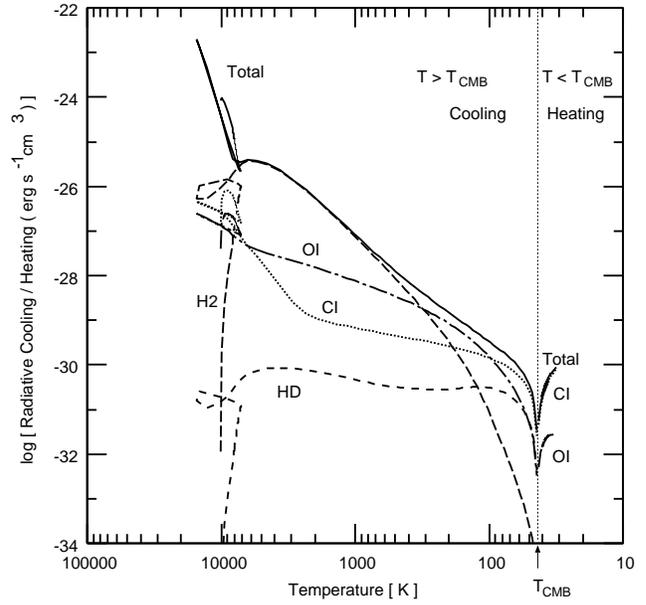}}
\end{center}
 \caption{Same as Fig. \ref{fig4}, 
 but for the fluid element (4) in Figure \ref{fig6}, 
 with metallicity $10^{-2} Z_{\odot}$.}
 \label{fig7}
\end{figure}

For a high metallicity gas, metal cooling or heating becomes important 
at low temperatures ($\lesssim 100$ K).
In this section, we elaborate on this effect on the evolution of the shell.
Fig. \ref{fig6} shows the thermal evolution of some 
fluid elements for the parameters $E_{\rm SN}=10^{52}$ ergs, $n_0=1$
cm$^{-3}$, and $Z=10^{-2} Z_{\odot}$, which is 100 times higher in
metallicity than in the fiducial case.  
The trajectories are almost the same as those in the fiducial run
until they reach $\sim 200$ K where the H$_2$ cooling rate sharply
decreases.  
In this case, the O{\sc i} and then C{\sc i} become the dominant 
coolants for $< 300$ K and $<100$K, respectively (Fig. \ref{fig7}).
Unlike in the fiducial case, the temperatures approach the CMB value 
through radiative cooling rather than through the adiabatic expansion.
With a rapid decrease in the radiative cooling rate 
near the CMB temperature, the shell begins to distend in a manner
similar to the fiducial case.    
Owing to the metal cooling, the temperatures at which 
the shell starts to swell are lower than those in the fiducial case. 
When the temperature falls below the CMB value, 
the radiative heating via C{\sc i} lines is effective 
enough to keep the temperature nearly constant at $\sim 40$ K, 
although with a gradual decrease in the temperatures. 
Thus, the final temperatures are somewhat higher than in the fiducial
case. 

How much metallicity is needed to affect the evolution of the shell?
To answer this question, we performed other simulations with 
metallicity $Z=10^{-3} Z_{\odot}$, and found that the metal cooling 
remains below that of HD molecules.
Hence, in this case, 
the evolution of the shell is almost the same as in the fiducial case. 
We conclude that the minimum metallicity necessary to affect the thermal
evolution of the shell is $\sim 10^{-2}Z_{\odot}$. 
We will mention its influence on the conditions 
for the shell fragmentation in Section \ref{sec:fragment}.


\subsection{Fragmentation of the Shell}
\label{sec:fragment}

\subsubsection{Overview of Linear Perturbation Analyses}

Here, we discuss whether the supernova-driven shell fragments into  
pieces before it dissolves into the ambient medium. 
When the shell fragments by gravitational instability, 
the next-generation stars are
expected to form from the fragments. 
Conditions for fragmentation of an expanding and decelerating shell have
been studied by several authors both analytically (Vishniac 1983;
Nishi 1992; Elmegreen 1994; Whitworth et al. 1994) and numerically
(Yoshida \& Habe 1992; Mac Low \& Norman 1993).  
The linear analysis by Elmegreen (1994) showed that the instantaneous
growth rate of density perturbations for transverse motion at wavenumber
$k$ is given by 
\begin{equation}
\omega (k,t) = - 3 \frac{V}{R} + \sqrt{ \frac{V^2}{R^2}   
- k^2 c_s^2 + 2 \pi G \Sigma_0 k}, \label{eq:grate}
\end{equation}
where R, V, $\Sigma_0$, and $c_s$ are the radius, velocity, 
unperturbed column density, and sound speed of the shell, 
respectively.
Here, the perturbations are assumed to grow in a way similar to
$\exp(\omega t)$.
The seeds of density perturbations are generated, for instance, 
by the decelerating shock instability (Mac Low \& Norman 1993), 
or by the shock propagation into the inhomogeneous medium.
The perturbations grow by collecting the ambient medium, while 
the new mass accretion onto the shell increases total transverse
momentum.  
In addition, the spherical expansion stretches and attenuates perturbations. 
These processes are taken into account via the $V/R$ terms in equation 
(\ref{eq:grate}) and the net effect acts to hinder 
growth of the perturbations. 
The maximum growth rate,  
\begin{equation}
\omega_{\rm max} =
 - 3 \frac{V}{R} + \sqrt{\left( \frac{V}{R}
						     \right)^2  
+ \left(\frac{\pi G \Sigma_0}{c_s} \right)^2}, 
\label{eq:maxrate}
\end{equation} 
is attained at wavenumber
\begin{equation}
 k_{\rm max} = \frac{\pi G \Sigma_0}{c_s^2}.
\label{eq:kmax}
\end{equation}
The shell becomes gravitationally unstable if $\omega_{\rm max} > 0$,
i.e., 
\begin{equation}
t_{\rm exp} \equiv \frac{R}{V} > \frac{8^{1/2} c_s}{\pi G \Sigma_0} \sim
 \frac{t_{\rm ff}^2}{t_{\rm cross}}, \label{eq:fragcon}
\end{equation}
where $t_{\rm exp}$, $t_{\rm ff}$, and $t_{\rm cross}$ are the
shell expansion time, free-fall time in the shell, 
and the sound crossing time for the shell width, respectively. 

Even when the shell becomes gravitationally unstable, 
it takes some time for the perturbations to grow to induce the
fragmentation. 
The time-scale for fragmentation can be roughly estimated by 
$1/\omega_{\rm max}$.
It should be noted, however, that the growth rate for a fixed wavenumber
changes with time because $V/R$ and $\Sigma_0$ in equation
(\ref{eq:grate}) are modified by the mass accretion.
To evaluate the typical fragmentation scale, 
we keep track of all the modes that attain the maximum growth rate
during the shell history.
For those modes, we define the fragmentation time $t_f$ as:  
\begin{equation}
\int_{t_i}^{t_f} \omega (k,t) dt = 1, 
\label{eq:tfrag}
\end{equation}
where $t_i$ is the time when the mode $k$ first becomes unstable.
Namely, $t_f$ is the time for the perturbation to grow $e$ times 
the initial value.
When condition (\ref{eq:tfrag}) is met, the shell is assumed to 
fragment into pieces with scales corresponding to  
$k_{\rm max}$ (Ehlerov\'{a} et al. 1997; Elmegreen, Palou\v{s} \&
Ehlerov\'{a} 2002).  
We evaluate the fragment mass by
\begin{equation}
M_{\rm frag} (\lambda_{\rm max}, t_f)
= \pi \left( \frac{\lambda_{\rm max}}{2} \right)^2 \Sigma_0 (t_f),
\label{eq:fmass}
\end{equation}
where $\lambda_{\rm max} \equiv 2 \pi/k_{\rm max}$.

The shell loses its identity when it slows down to
the sound speed of the ambient medium. 
For triggering subsequent star formation,
fragmentation must occur within the lifetime of the shell.

\subsubsection{Application of the Linear Analysis}

In this section, we use the result of the linear analysis
to find the conditions for fragmentation of a supernova-driven shell.
The dispersion relation (\ref{eq:maxrate}) is derived with 
one-zone approximation to the shell, whereas the shell has a
stratified structure in our simulations. 
The application of the linear analysis to our result is, therefore, 
not straightforward.
If the stratified structure is maintained until the shell disappears, 
the gravitational instability seems to occur first  
in a cold layer of the shell due to its low temperature.
However, this is not the case because the mass of 
the cooled gas is not sufficient to trigger 
the gravitational instability.
On the contrary, the coldest layer tends to gradually expand 
as discussed in Section \ref{result2}, 
indicating that the self-gravity is not important there. 
In reality, the gravitational instability most 
easily occurs at the scale comparable to the width of the shell 
because the column density of the layer must be high 
enough for fragmentation.
We then simply estimate a representative temperature 
by taking mass-weighted average over the whole shell.  

\begin{figure}
\begin{center}
 \scalebox{1.0}{
  \includegraphics*[width=\hsize]{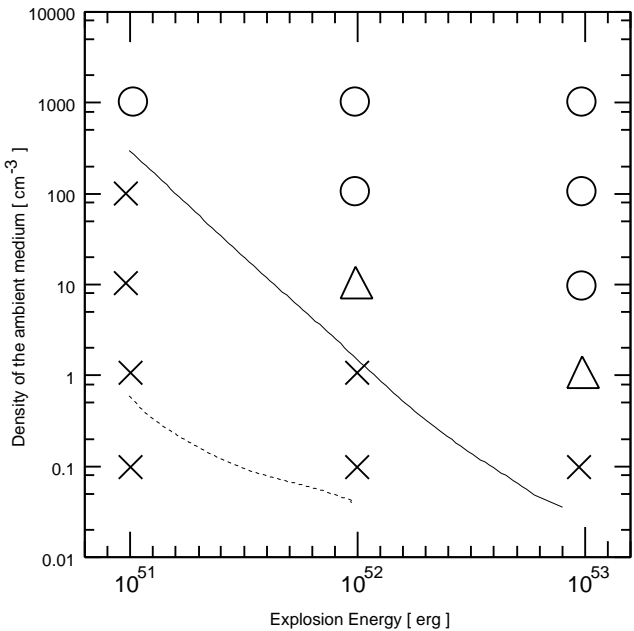}}
\end{center}
 \caption{Dependence of the fragmentation result of the shell on 
  the ambient density and explosion energy for metallicities
 $Z=10^{-2} Z_{\odot}$ and $Z=10^{-4} Z_{\odot}$.
 In the cases of $E_{\rm SN}=10^{51}$ and $10^{52}$ erg, the circles,
 triangle, and crosses mean that the fragmentation occurs 
 for both metallicities, only for $Z=10^{-2} Z_{\odot}$, and for
 neither, respectively.
 In the cases of $E_{\rm SN}=10^{53}$ erg, the circles and crosses
 represent the successes and fails of the fragmentaion respectively, and
 the triangle satisfies the condition (\ref{eq:fragcon})
 but does not equation (\ref{eq:tfrag}). 
 Also plotted are the results by Salvaterra et al. (2004; solid) and Machida 
 et al. (2005; dotted), above which the fragmentation occurs.}  
 \label{fig8}
\end{figure}

In Fig. \ref{fig8}, we show 
the parameter range of ($E_{\rm SN}, n_0)$ 
where the shell fragments within its lifetime
for runs with $Z=10^{-4} Z_{\odot}$ and $10^{-2} Z_{\odot}$.
Metallicity affects the result only for $(E_{\rm SN},n_0)=(10^{52}, 10)$, 
where the $10^{-4} Z_{\odot}$ run satisfies condition 
(\ref{eq:fragcon}) but not (\ref{eq:tfrag}) 
before the dissolution of the shell.
For high ambient density, the shell accumulates 
a large amount of gas within the lifetime and fragments even in the case
of $10^{-4} Z_{\odot}$.  
Thus, the metallicity dependence disappears for such high ambient densities.
In the cases of $E_{\rm SN}=10^{52}{\rm erg}$, 
the shell fragmentation occurs for both metallicities 
at the ambient density $\geqq 10^2$ cm$^{-3}$. 
For lower explosion energy $E_{\rm SN}=10^{51}{\rm erg}$, 
this threshold density increases 
to $10^{3}{\rm cm^{-3}}$. 
In Fig. \ref{fig8}, we also plot the conditions for shell 
fragmentation found by Salvaterra et al. (2004) and 
Machida et al. (2005) for comparison.
The discrepancy among their conditions stems from the difference in
their adopted shell temperatures and criteria for fragmentation.
Our result almost coincides with that by Salvaterra et al. (2004)
because their assumed temperature is almost the same as ours.
As long as the mean temperature falls to $200$ -- $300$ K 
through H$_2$ cooling, the total amount of 
gas swept up into the shell does matter for fragmentation.
Thus, the conditions for fragmentation rely heavily on 
the ambient density as well as explosion energy, 
but little on the metallicity.

In Table \ref{tab:timef}, we show the properties of the shell 
at fragmentation time $t_f$. 
There is a range of the fragment mass because some unstable modes
fulfill the condition (\ref{eq:tfrag}) soon after $t_f$.
This means that the fragmentation scale depends largely 
on the shape of initial perturbation, or that the shell fragments with a
wide range of mass scales.
The fragment mass for models with $n_0 \geqq 10^2$ cm$^{-3}$ 
is in the range $10^{2}$ -- $10^{3}$ M$_{\odot}$, while that for 
a model with $n_0=10$ cm$^{-3}$ is an order of magnitude lower. 
Because the wavelength $\lambda_{\rm max}$ lengthens 
for the smaller column density (see equation \ref{eq:kmax}),
the fragment mass becomes higher for lower ambient density.

\begin{table*}
\begin{center}
\scalebox{0.85}{
\begin{tabular}{cccccccccccccc}
\hline \hline
 ($E_{\rm SN},n_0,Z$) & $T_{\rm sh}$ & $\Sigma_0$ 
 & $\Delta R$ & $\bar{\rho}_{\rm sh}$ 
 & $t_{\rm max}$ & $t_f$ & $\lambda_{\rm max}$ 
 & $R_{\rm sh}$ & $M_{\rm sh}$ 
 & $M_{\rm frag}$ & $M_{\rm frag}^{\rm min}$ & $M_{\rm frag}^{\rm max}$
 & $M_{\rm crit}$ \\
 (erg, cm$^{-3}$, $Z_{\odot}$) & (K) & (g/cm$^{-2}$)
 & (pc) & (g/cm$^{-3}$)
 & (Myr) & (Myr) & (pc) 
 & (pc) & (M$_{\odot}$)
 & (M$_{\odot}$) & (M$_{\odot}$) & (M$_{\odot}$) & (M$_{\odot}$) \\
\hline 
($10^{51},10^3,10^{-4}$) & $46$ & $2.2 \times 10^{-2}$ & $0.78$ 
 & $9.2 \times 10^{-21}$ & $1.0$ & $1.7$ & $3.6$  
 & $9.9$ & $1.3 \times 10^5$
 & $1.1 \times 10^3$ & $2.8 \times 10^2$  & $2.7 \times 10^3$ 
 & $6.6 \times 10^7$ 
\\
($10^{51},10^3,10^{-2}$) & $45$ & $2.3 \times 10^{-2}$ & $0.74$ 
 & $1.0 \times 10^{-20}$ & $0.53$ & $1.6$ & $3.1$  
 & $10$ & $1.3 \times 10^5$
 & $8.1 \times 10^2$ & $2.4 \times 10^2$  & $8.5 \times 10^2$ 
 & $6.7 \times 10^7$ 
\\
($10^{52},10^3,10^{-4}$) & $83$ & $3.7 \times 10^{-2}$ & $0.54$  
 & $2.3 \times 10^{-20}$ & $0.84$ & $1.4$ & $2.6$ & $17$  
 & $6.6 \times 10^5$ & $9.5 \times 10^2$
 & $5.9 \times 10^2$  & $2.2 \times 10^3$ 
 & $3.3 \times 10^8$ 
\\
($10^{52},10^3,10^{-2}$) & $46$ & $3.7 \times 10^{-2}$ & $0.49$  
 & $2.4 \times 10^{-20}$ & $0.35$ & $1.5$ & $2.0$ & $16$  
 & $ 5.5 \times 10^5$ & $5.8 \times 10^2$
 & $1.4 \times 10^2$  & $6.1 \times 10^2$ 
 & $2.7 \times 10^8$ 
\\
($10^{52},10^2,10^{-4}$) & $75$ & $9.8 \times 10^{-3}$ & $2.1$ 
 & $1.6 \times 10^{-21}$ & $3.0$ & $4.5$ & $9.3$ 
 & $43$ & $1.1 \times 10^6$
 & $3.2 \times 10^3$
 & $2.3 \times 10^2$  & $5.8 \times 10^3$ 
 & $1.8 \times 10^8$ 
\\
($10^{52},10^2,10^{-2}$) & $51$ & $9.4 \times 10^{-3}$ & $1.6$  
 & $1.9 \times 10^{-21}$ & $2.4$ & $4.0$ & $7.0$ & $41$  
 & $9.7 \times 10^5$ & $1.7 \times 10^3$
 & $3.9 \times 10^2$  & $2.3 \times 10^3$ 
 & $1.5 \times 10^8$ 
\\
($10^{52},10,10^{-2}$) & $55$ & $2.5 \times 10^{-3}$ & $7.4$ 
 & $1.2 \times 10^{-22}$ & $9.3$ & $15$ & $29$
 & $110$ & $1.9 \times 10^6$
 & $8.0 \times 10^3$
 & $4.4 \times 10^3$  & $2.0 \times 10^4$ 
 & $9.6 \times 10^7$ 
\\
($10^{53},10^3,10^{-4}$) & $87$ & $5.5 \times 10^{-2}$ & $0.16$ 
 & $1.1 \times 10^{-19}$ & $0.33$ & $0.88$ & $2.3$ 
 & $24$ & $1.9 \times 10^6$
 & $1.1 \times 10^3$
 & $1.9 \times 10^2$  & $1.5 \times 10^3$ 
 & $9.2 \times 10^8$ 
\\
($10^{53},10^2,10^{-4}$) & $78$ & $1.5 \times 10^{-2}$ & $0.61$ 
 & $7.8 \times 10^{-20}$ & $2.2$ & $3.0$ & $5.7$ 
 & $65$ & $3.7 \times 10^6$
 & $1.8 \times 10^3$
 & $6.1 \times 10^2$  & $3.3 \times 10^3$ 
 & $5.8 \times 10^8$ 
\\
($10^{53},10,10^{-4}$) & $105$ & $4.1 \times 10^{-2}$ & $3.7$ 
 & $3.6 \times 10^{-22}$ & $6.9$ & $11$ & $26$ 
 & $180$ & $7.9 \times 10^6$
 & $9.9 \times 10^3$
 & $8.5 \times 10^3$  & $1.5 \times 10^4$ 
 & $1.8 \times 10^7$ 
\\
\hline \hline
\\
\end{tabular}
}
\end{center}
\caption{Properties of the shell when the shell fragmentation first
 occurs. The left column is the model parameters. 
 The notation $T_{\rm sh}$ represents the mass-weighted mean temperature
 in the shell, $\Sigma_0$ the column
 density, $\Delta R$ the width of the shell, $\bar{\rho}_{\rm sh}$ the
 mean density in the shell, $t_{\rm max}$ the time when the shell 
 becomes gravitationally unstable first, $t_f$ the fragmentation time,  
 $\lambda_{\rm max}$ the wavelength of perturbation which becomes 
 unstable first, $R_{\rm sh}$ and $M_{\rm sh}$ the radius and mass of
 the shell at fragmentation, and $M_{\rm frag}$ the mass of the
 fragments, $M_{\rm frag}^{\rm min}$ the minimum mass of the fragments
 after $t_{\rm f}$, $M_{\rm frag}^{\rm max}$ the maximum mass of the
 fragments after $t_{\rm f}$. The last column is the critical halo mass
 below which star formation cannot occur in the induced manner.} 
\label{tab:timef}
\end{table*}
  
\section{Discussions}
\label{sec:discuss}

\subsection{Significance of Star Formation Triggered by Supernova Explosions}

The supernova explosion is an important feedback process 
influencing cosmic star formation history as well as 
the nearby star formation activity.
We have shown that supernova explosions induce subsequent star formation 
as a result of shell fragmentation in a high-density environment 
(see Fig. \ref{fig8}).
Here, we estimate the critical mass of haloes above which
second-generation stars form in the induced manner, using our results. 
We assume that the explosion occurs at the centre of the halo. 
We have derived the minimum ambient density $\rho_{\rm min}$
for a given explosion energy, at which the shell fragments 
at a radius $r_{\rm f}$.
Thus, the condition for the shell fragmentation
is that the column density exceeds 
$\rho_{\rm min} r_{\rm f}/3$ at a radius $< r_{\rm f}$, or in other
words, the mean density $\bar{\rho}_{\rm f}$ within the 
radius $r_{\rm f}$ is lower than $\rho_{\rm min}$.
We evaluate the mean density $\bar{\rho}_{\rm f}$ 
assuming the density profile in the halo 
$\rho(r) \propto r^{-w}$, with the index $w \simeq 2$ 
as suggested by numerical calculations of the first star formation.

Using the mean density $\bar{\rho}_{\rm vir}(z_c)$ and 
virial radius $r_{\rm vir}$ of a halo viliarizing at $z_c$, 
we obtain the mean baryonic density inside $r_{\rm f}$ as 
\begin{align}
\bar{\rho}_{\rm f} &= \frac{\Omega_b}{\Omega_m}
\bar{\rho}_{\rm vir} (z_c)
\left( \frac{r_{\rm f}}{r_{\rm vir}} \right)^{-2} \label{eq:mrhof} \\
& \simeq 3.0 \times 10^{-25} ~{\rm g~cm}^{-3} 
\left( \frac{1+z_c}{16} \right)^{3} \left( \frac{r_{\rm f}}{r_{\rm vir}}
				    \right)^{-2}.
\end{align} 
Here, we apply the spherical collapse model to the mean density
in the halo, $\bar{\rho}_{\rm vir}=18 \pi^2 (\Omega_b/\Omega_m) \rho_b(z_c)$, 
where $\rho_b$ is the background density (e.g., Padmanabhan 1993). 
By eliminating $r_{\rm vir}$ in equation (\ref{eq:mrhof}) with 
$M_{\rm halo} = (4 \pi / 3) \bar{\rho}_{\rm vir} r_{\rm vir}^3$, 
the halo mass $M_{\rm halo}$ can be expressed as 
\begin{equation}
M_{\rm halo} = \frac{4 \pi}{3} \bar{\rho}_{\rm vir}^{-1/2} (z_c) r_{\rm
 f}^3 \left( \frac{\Omega_b}{\Omega_m} \right)^{-3/2}
 \bar{\rho}^{3/2}_{\rm f}.
\end{equation}
Setting $\bar{\rho}_{\rm f} = \rho_{\rm min}$, we obtain the
critical halo mass for triggered star formation: 
\begin{equation}
M_{\rm halo}^{\rm crit} \simeq 1.5 \times 10^{8} {\rm M}_{\odot} 
\left( \frac{1+z_c}{16} \right)^{-1.5} 
\left( \frac{r_{\rm f}}{41.4 {\rm pc}} \right)^{3}
\left( \frac{n_{\rm min}}{10^2 {\rm cm}^{-3}} \right)^{1.5},
\label{eq:Mhalo}
\end{equation}
where $n_{\rm min}=\rho_{\rm min}/\mu m_{\rm H}$.
Note that the dependence on ambient density $n_{\rm min}$ 
is also included through the fragmentation radius $r_{\rm f}$.
The critical halo masses range from $1.8 \times 10^7$ M$_{\odot}$ to
$9.2 \times 10^8$ M$_{\odot}$. These values are listed in 
Table \ref{tab:timef}.
These mass-scales correspond to $\approx 3 \sigma$ fluctuations 
virializing at $z=15$ and the abundance of such objects 
is about a few per cubic Mpc. 
The critical mass for triggered star formation $M_{\rm halo}^{\rm crit}$
exceeds the halo mass of $4.7 \times 10^7$ M$_{\odot}$
$[(1+z_c)/16]^{-3/2}$ corresponding to the virial temperature of $10^4$
K(Barkana \& Loeb 2001), and consequently, star formation triggered by a
supernova explosion is unlikely to occur in low-mass haloes 
($T_{\rm vir} < 10^4$ K) where primordial stars form solely via H$_2$ cooling. 
Instead, in such low-mass haloes, the expanding SNR evacuates 
all the gas and quenches the subsequent star formation activity.
The triggering process becomes increasingly important at lower 
redshift as more massive haloes of $> 10^{8-9} M_\odot$ begin to form.

So far, we have assumed the density distribution $\rho \propto r^{-2}$, 
which is appropriate at the stellar birth.
If an expanding H{\sc ii} region created during the stellar 
lifetime disturbs and even blows the surrounding material away from
the host halo before the supernova explosion, the column density of the
supernova-driven shell is significantly reduced and the subsequent
triggering process does not work.
This effect is important in low-mass ($\lesssim 10^6 M_\odot$) 
haloes (Kitayama \& Yoshida 2005; Whalen et al. 2008).
However, in massive haloes ($\gtrsim 10^8 M_\odot$) where 
the triggered process is possible, 
the H{\sc ii} region is trapped deep inside the virial radius 
and barely removes any gas from the host haloes.
Thus, the expansion of the H{\sc ii} region does not change our criteria
for the triggered star formation.
Here, it should be noted that the densities in the centre of the trapped
H{\sc ii} regions are $10^4$ -- $10^7$ cm$^{-3}$ and much higher than we
have examined, according to some numerical simulations (Kitayama \&
Yoshida 2005; Whalen et al. 2008). 
The reason why we have investigated the density range of $10^2$ --
$10^3$ cm$^{-3}$ is only for comparison with the previous works of
Salvaterra et al. (2004) and Machida et al. (2005).

Whalen et al. (2008) studied supernova explosions in 
neutral haloes undisturbed by the expansion of H{\sc ii} regions.
They showed that the SNR is trapped by the deep gravitational
potential well of the dark matter. 
It initially expands out to $\sim 50$ pc, and then turns to 
fall back. Metals in the ejecta will mix with gas in the halo 
before the SNR stagnates. The polluted gas may fragment into clumps
through metal fine-structure line cooling, which leads to subsequent
star formation. This triggering mode is beyond the scope of this paper, 
and to be examined in future studies.
In addition, we have ignored another process of possibly triggering star
formation described below because of a limitation of our simulations.
The H{\sc ii} region prior to the supernova sweeps up materials into a
shell (Kitayama et al. 2004; Whalen et al. 2004). 
The SNR catches up with the pre-existing shell.
Whalen et al. (2008) showed that the collision is quite violent
and would mix heavy elements from the ejecta with materials
in the shell. 
Fracturing the ejecta into the dense shell and the resultant enhanced
cooling could lead to star formation. 
The shell/ejecta interaction occurs even in low-mass halos of 
$\lesssim 10^8 M_\odot$.  
Thus, simulations with more realistic density distribution
are needed for a complete survey of the conditions required for 
triggered star formation.

We finally note that hydrodynamical mixing of metals in the SNR
is generally important for realistic evolution. 
Even if the ambient medium is constituted by pristine H and He gas, the
polluted gas will cool via metal fine-structure lines.
Although our calculations with pre-enriched ambient medium
predict that the critical halo mass needed for triggering star formation
depends hardly on metallicity, it is still uncertain how the 
hydrodynamical mixing spreads metals into the SNR and affects 
the triggering processes.

\subsection{Subsequent Evolution of the Fragments 
and Possibility of Low-mass Star Formation}
\label{sec:discuss2}

After the fragmentation, each fragment continues to contract 
through its own gravity.
Although the mass of fragments is $\sim 10^3$ -- $10^4~M_\odot$ 
(Table \ref{tab:timef}), they are expected to fragment once again into
smaller ($< 10~M_\odot$) pieces. 
The shell initially fragments into disk-like clouds, 
and then further fragmentation leads to smaller 
filamentary clouds (Miyama, Narita \& Hayashi 1987).
Nakamura \& Umemura (2001, 2002) examined fragmentation 
of primordial filamentary clouds and concluded that 
they fragment into spherical cores of $1$ -- $10^2$ M$_{\odot}$, 
the exact value of which depends on the initial cloud density $n_c$ and 
gravity-to-pressure ratio $f$.
The fragmentation leads to low-mass ($\sim 1M_{\odot}$) cores 
only if the conditions that $n_c \gtrsim 10^5$ cm$^{-3}$ and $f>3$ are met.
Salvaterra et al. (2004) pointed out that these conditions are easily 
satisfied and thus low-mass star formation occurs
in the supernova-driven shell in primordial environments. 
However, our more detailed calculations 
demonstrated that the number density in the shell never 
exceeds $10^{5}$ cm$^{-3}$ for the ambient density 
$n_0 \leq 10^3~{\rm cm}^{-3}$, and the low-mass stars 
are unlikely to form without any metal enrichment 
even in the triggered fashion.

Provided that the metallicity is higher than 
$Z_{\rm crit}=10^{-5}$ -- $10^{-6} Z_{\odot}$ and 
a sizeable portion $\sim 0.5-1$ of those metals 
condenses into dust grains, fragmentation is induced by dust 
cooling at high densities $>10^{10}$ cm$^{-3}$ 
(Schneider et al. 2002, 2006; Omukai et al. 2005; Tsuribe \& Omukai 2006,
2008; Clark, Glover \& Klessen 2008).
In the early universe, dust grains originate from 
supernovae (Todini \& Ferrara 2002; Nozawa et al. 2003; 
Schneider, Ferrara \& Salvaterra 2004) rather than the AGB stars.
Nozawa et al. (2007) studied the formation of dust grains in 
population III supernovae and their destruction by the reverse shock.
They concluded that the mass of the surviving dust grains in the shell is at
most $0.1$ M$_{\odot}$ for the ambient density $n_0=10$ cm$^{-3}$ 
at $t \sim 0.7$ Myr. 
To reach the critical metallicity $Z_{\rm cr}$, 
the total mass in the shell should be less than
$5 \times 10^6 ~(Z_{\rm cr}/10^{-6} Z_{\odot})^{-1} ~\rm{M}_{\odot}$ 
before fragmentation, under the assumption of homogeneous 
grain distribution within the shell.
Because the shell mass at fragmentation 
is $\sim 10^{6}$ M$_{\odot}$ for $n_0=10$ cm$^{-3}$ 
(Sec.\ref{sec:results}),
low-mass stars can be triggered by the first supernova.

\subsection{Effects of External Photo-dissociating Radiation}

We assume that H$_2$ molecules in the ambient medium are completely
photo-dissociated by the central star before the supernova, but they are
replenished quickly due to a high fraction of the catalyst electrons in
our calculations. 
As a result, the ambient temperature decreases to several hundreds of
Kelvins owing to the H$_2$ cooling (Fig. \ref{fig1}).
However, we have not considered the effects of any external radiation.
Once stars appear in the universe, they build up a background of H$_2$
dissociating radiation, which delays subsequent star formation  
(e.g., Haiman, Rees \& Loeb 1997; Haiman, Abel \& Rees 2000).
If an external radiation flux dissociates the H$_2$ molecules, the gas
temperature remains higher because of inefficient cooling. 
The low temperature in the ambient medium prolongs the lifetime of the
shell. The fragmentation time is slightly shorter than the shell lifetime
even in our calculation, and thus the fragmentation may not occur with
external heating.  

According to a cosmological simulation by Johnson, Greif \& Bromm (2008), 
the photo-dissociating radiation flux is $J_{21} \lesssim 0.04$ 
at $12.87$ eV in units of $10^{-21}$ erg s$^{-1}$ cm$^{-2}$ 
Hz$^{-1}$ sr$^{-1}$.   
For this background, the H$_2$ dissociation timescale is  
$t_{\rm diss} \approx 6 \times 10^5 (J_{21}/0.04)^{-1}$ yr 		      
(Abel et al. 1997), whereas the typical H$_2$ formation time behind the
shock in our calculation is $t_{\rm form} \sim 10^4$ yr 
at the beginning of the PDS phase.
The formation time is thus an order of magnitude shorter than
the dissociation time. 
Hence, the external photo-dissociating flux does not prevent the H$_2$
formation at this phase.
Soon after that, the H$_2$ column density of the shell 
rapidly increases to shield the photo-dissociation 
(i.e., $\gtrsim 10^{14}$ cm$^{-2}$;  
e.g., Draine \& Beltoldi 1996) as seen below. 
Before the photo-dissociation,
the H$_2$ column density of the shell reaches 
$\sim y({{\rm H}_2}) n_{\rm H} v_{\rm sh} t_{\rm diss}$, 
where $ y({{\rm H}_2})$ and $v_{\rm sh}$ are the
fractional abundance of H$_2$ molecules and shock velocity. 
The shock velocity is on the order of $10^{2}$ km s$^{-1}$ at the
incipient PDS phase, while $n_{\rm H}$ is enhanced to 
$\sim \mathcal{M}^2 n_0$ by an isothermal shock with the Mach number 
$\mathcal{M}$.
The H$_2$ fract
ion necessary for self-shielding is then only 
\begin{equation}
{\rm y}_{{\rm H}_2} \sim 10^{-8} \left( \frac{v_{\rm sh}}{10^2 ~{\rm km/s}}
			  \right)^{-1}
\left( \frac{t_{\rm diss}}{10^6 ~{\rm yr}} \right)^{-1}
\left( \frac{n_0}{1 ~{\rm cm}^{-3}} \right)^{-1},
\end{equation}
where we assume that $\mathcal{M} \sim 10$.
Because the H$_2$ abundance far exceeds this value within 
the dissociation time (Fig. \ref{fig5}), 
we conclude that the external photo-dissociating radiation does not
significantly prevent H$_2$ formation in the shell.

\subsection{Effects of Non-gravitational Instabilities}

In our analysis of gravitational instability, we have implicitly 
assumed that the coherence of the layer
is not destroyed by some dynamical instabilities
emerging before the gravitational one. 
For instance, the decelerating shock instability 
is expected to grow from the linear analysis (Vishniac 1983). 
Mac Low \& Norman (1993) numerically demonstrated, however, that 
this instability saturates at a rather low amplitude 
in the non-linear stage, 
and does not significantly disturb the shell structure.
The thermal instability is also expected to occur
in the rapid cooling phase in the shocked layer.
Koyama \& Inutsuka (2002) studied propagation of a shock into the
contemporary ISM and fragmentation of a shock-compressed layer, taking 
account of cooling, heating, and thermal conduction.
They showed that the thermal instability causes the cooled layer 
to fragment into small ($\sim$ $10$ -- $100$ AU) cloudlets with some
translational motions. 
Bromm, Yoshida \& Hernquist (2003) simulated the first supernova explosions, 
and showed that the swept-up shell consists of 
small cloudlets, which are presumably formed by thermal instability.
The motions of the cloudlets effectively enhance
the sound speed within the shell, leading to delay in growth
of the gravitational instability.
After the shell fragments through the gravitational instability,
the cloudlets will aggregate into a larger cloud 
with the contraction of the fragments.
This may ultimately lead to cluster star formation (Elmegreen 1998).
Thus, thermal instability influences not only the shell fragmentation
but also the subsequent star formation in the fragments.
Although our one-dimensional calculations cannot clarify the detailed
fragmentation processes in the shell and subsequent evolution of the
fragments, future multi-dimensional simulations are expected to 
reveal them.

\section{Summary}
\label{sec:summary}

To explore the possibility of star formation triggered by 
supernovae in the early universe, we have studied 
the evolution of shells formed around the supernova bubble 
during the snowplough phase in low-metallicity 
environments. 
We investigated detailed structure of the shell, using spherical
symmetric hydrodynamics with non-equilibrium chemistry for different
sets of ambient densities, explosion energies, and metallicities.
Our results are summarized as follows:

\begin{itemize}
\item[(i)] We have demarcated the ranges of metallicity
in terms of thermal evolution of the shell:
\begin{itemize}
\item
For metallicities with $< 10^{-2} Z_{\odot}$, the post-shock
gas first cools promptly to $\sim 200$ K through the H$_2$ cooling, 
and then progressively to a lower value through adiabatic expansion.
Although $10$ \% of the deuterium transforms into HD
in the shell, the HD cooling and heating have never become 
important because of decreasing density. 

\item
For higher metallicities ($> 10^{-2} Z_{\odot}$),
the fine-structure line transitions of O and C
reduce the post-shock gas below $100$ K.
After the radiative cooling becomes inefficient, 
the gas then adiabatically expands as in 
the lower-metallicity cases.
The CMB heating via the C{\sc i} fine-structure transitions
hinders the gas from cooling below the CMB temperature.
\end{itemize}

\item[(ii)] 
We have examined whether the swept-up shell becomes gravitationally
unstable before it mixes with the ISM,
using the linear perturbation analysis.
Fragmentation of the shell easily occurs in denser ISM and for higher
	   explosion energies because these conditions 
	   are favourable to attaining the column density
	   of the shell necessary for 
	   fragmentation.  
The criteria for the fragmentation depend little on metallicity as
	   thermal evolution of the shell is not so sensitive to it. 

\item[(iii)] 
We have evaluated the fragment mass from the wavelength of
the maximum growth modes.
Because the wavelength is inversely proportional to the column
density, the fragment mass becomes lower for both higher ambient densities
	   and higher explosion energies.
The resulting fragment mass is $10^2$ -- $10^3$ M$_{\odot}$.

\item[(iv)] 
We have derived the critical halo mass 
$M^{\rm crit}_{\rm halo} \simeq 10^8~M_\odot$ above which star formation 
can be triggered by a supernova explosion.
In minihaloes ($< 10^6 M_\odot$) where the 
first stars are more likely to form, this triggering mechanism does not
	   work. 

\end{itemize}

\section*{Acknowledgments}

We thank M. N. Machida, M. Nagashima, and S. Inutsuka for helpful
comments and discussions.
This study is supported in part by Research Fellowships of the Japan
Society for the Promotion of Science for Young Scientists 
(T. N. and T. H.).

\appendix

\section{Chemical Reactions}

We list the chemical reactions we include in
Table \ref{tab:reactions}. 

The definitive version is available at 'www.blackwellsynergy.com'.

\begin{table}
\setlength{\tabcolsep}{5pt}
\renewcommand{\arraystretch}{0.3}
\begin{center}
\scalebox{1.0}{
\begin{tabular}{llc}
\hline \hline
\\
Number & Reactions & References \\
\\
\hline \\
R1&  H \ +\ e$^{-}$ $\rightarrow$ H$^+$ \ +\ 2 e$^{-}$ & 1 \\
\\
R2&  H$^+$  + e$^{-}$ $\rightarrow$ H + $\gamma$  &  1 \\
\\
R3&  He + e$^{-}$ $\rightarrow$ He$^+$ + 2 e$^{-}$ & 1 \\
\\
R4&He$^{+}$ + e$^{-}$ $\rightarrow$ He + $\gamma$ & 2 \\
\\
R5&  He$^+$ + e$^{-}$ $\rightarrow$ He$^{++}$ + 2e$^{-}$ & 1 \\
\\
R6&  He$^{++}$ + e$^{-}$ $\rightarrow$ He$^{+}$ + $\gamma$ & 2 \\
\\
R7&  H  + e$^{-}$ $\rightarrow$ H$^-$ + $\gamma$ & 2 \\
\\
R8&  H + H$^-$ $\rightarrow$ H$_2$ + e$^{-}$ & 2 \\ 
\\
R9& H   +  H$^{+}$   $\rightarrow$  H$_{2}^{+}$   +  $\gamma$ & 2 \\
\\
R10& H$_{2}^{+}$   +  H   $\rightarrow$  H$_2$*   +  H$^{+}$ & 2\\
\\
R11& H$_2$ +H$^+$ $\rightarrow$ H$_2^{+}$ + H & 2\\ 
\\
R12& H$_2$ +\ e$^{-}$ $\rightarrow$ 2H + e$^{-}$     &  1 \\
\\
R13 &  H$_2$   +  H   $\rightarrow$  3H &  3 \\
\\
R14 & H$^-$ + e$^{-}$ $\rightarrow$ H + 2e$^{-}$ & 1 \\
\\
R15&  H$^{-}$   +  H   $\rightarrow$  2H   +  e$^{-}$  & 1 \\
\\
R16& H$^{-}$ + H$^{+}$ $\rightarrow$  2H & 2 \\
\\
R17&  H$^{-}$   +  H$^{+}$   $\rightarrow$  H$_{2}^{+}$   +  e$^{-}$ & 2\\ 
\\
R18 &  H$_{2}^{+}$   +  e$^{-}$   $\rightarrow$  2H & 2 \\
\\
R19 &  H$_{2}^{+}$   +  H$^{-}$   $\rightarrow$  H   +  H$_2$ & 1 \\
\\
R20 &  D$^{+}$   +  e$^{-}$   $\rightarrow$  D   +  $\gamma$ & 4 \\
\\
R21 &  D   +  H$^{+}$   $\rightarrow$  D$^{+}$   +  H & 4 \\
\\
R22 &  D$^{+}$   +  H   $\rightarrow$  D   +  H$^{+}$ & 4 \\
\\
R23 &  D   +  H   $\rightarrow$  HD   +  $\gamma$ & 4 \\
\\
R24 &  D   +  H$_2$   $\rightarrow$  H   +  HD & 4 \\
\\
R25 &  HD$^{+}$   +  H   $\rightarrow$  H$^{+}$   +  HD & 4 \\
\\
R26 &  D$^{+}$   +  H$_2$   $\rightarrow$  H$^{+}$   +  HD & 5 \\
\\
R27 &  HD   +  H   $\rightarrow$  H$_2$   +  D & 4 \\
\\
R28 &  HD   +  H$^{+}$   $\rightarrow$  H$_2$   +  D$^{+}$ & 5 \\
\\
R29 &  D   +  H$^{+}$   $\rightarrow$  HD$^{+}$   +  $\gamma$ & 4 \\
\\
R30 &  D$^{+}$   +  H   $\rightarrow$  HD$^{+}$   +  $\gamma$ & 4 \\
\\
R31 &  HD$^{+}$   +  e$^{-}$   $\rightarrow$  H   +  D & 4 \\
\\
R32 &  D   +  e$^{-}$   $\rightarrow$  D$^{-}$   +  $\gamma$ & 2 \\
\\
R33 &  D$^{+}$   +  D$^{-}$   $\rightarrow$  2D & 2 \\
\\
R34 &  H$^{+}$   +  D$^{-}$   $\rightarrow$  D   +  H & 2 \\
\\
R35 &  H$^{-}$   +  D   $\rightarrow$  H   +  D$^{-}$ & 2 \\
\\
R36 &  D$^{-}$   +  H   $\rightarrow$  D   +  H$^{-}$ & 2 \\
\\
R37 &  D$^{-}$   +  H   $\rightarrow$  HD   +  e$^{-}$ & 2 \\
\\
R38 &  H  +  H + H  $\rightarrow$  H$_2$ + H & 6,7 \\
\\
R39 &  H  +  H  + H$_2$  $\rightarrow$  H$_2$   +  H$_2$ & 6 \\
\\
R40 & H$_2$  +  H$_2$  $\rightarrow$   H  +  H  +  H$_2$  & 3,6 \\
\\
R41 & H  +  H  $\rightarrow$  H$^{+}$   +  e$^{-}$   +  H & 6 \\
\\
R42 & C$^{+}$  +  e$^{-}$  $\rightarrow$  C  +  $\gamma$ & 7 \\
\\
R43 &  O$^{+}$  +  e$^{-}$  $\rightarrow$  O  +  $\gamma$  & 8 \\ 
\\
R44 & O$^{+}$  +  H   $\rightarrow$  H$^{+}$  +  O & 7 \\
\\
R45 &  O   +  H$^{+}$  $\rightarrow$  O$^{+}$  +  H  & 7 \\
\\
R46 &  He  +  H$^{+}$  $\rightarrow$  He$^{+}$  +  H & 2 \\
\\
R47 & He$^{+}$  +  H  $\rightarrow$  He  +  H$^{+}$ & 2 \\
\\
R48 & He  +  H$^{+}$  $\rightarrow$  HeH$^{+}$  + $\gamma$ & 2 \\
\\
R49 &  He  +  H$^{+}$  $\rightarrow$  HeH$^{+}$  +  $\gamma$ & 2 \\
\\
R50 &  He  +  H$_{2}^{+}$  $\rightarrow$  HeH$^{+}$  +  H & 2 \\
\\
R51 &  He$^{+}$  +  H  $\rightarrow$  HeH$^{+}$  +  $\gamma$ & 2 \\
\\
R52 &  HeH$^{+}$  +  H  $\rightarrow$  He  +  H$_{2}^{+}$ & 2 \\
\\
R53 &  HeH$^{+}$  +  e$^{-}$  $\rightarrow$  He  +  H & 2 \\
\\
\hline \hline
\\
\end{tabular}
}
\end{center}
\caption{
References.---1. Abel et al. (1997); 2. Galli \& Palla (1998); 3. Shapiro
 \& Kang (1987); 4. Stancil, Lepp, \& Dalgarno (1998); 5. Galli \& Palla
 (2002); 6. Palla, Salpeter \& Stahler (1983); 7. Stancil et al. (1999); 
 8. Nahar \& Pradhan (1997).}
\label{tab:reactions}
\end{table}

\end{document}